\documentstyle[preprint,aps,eqsecnum,epsf,floats]{revtex}
\tighten
\begin{document}
\preprint{NT@UW-98-27}
\draft
\title{The Quark Distributions in the $\Sigma^+$ Hyperon}
\author{Mary Alberg$^{a,b,}$\footnote{E-mail: alberg@phys.washington.edu},
 Thomas Falter$^{a,c,}$\footnote{E-mail: Thomas.Falter@physik.uni-giessen.de}
 and Ernest M. Henley$^{a,d,}$\footnote{E-mail: henley@phys.washington.edu}}
\address{{\it $^a)$ Department of Physics, Box 351560,
University of Washington,
Seattle, WA 98195, USA}\\
{\it $^b)$ Department of Physics,
Seattle University,
Seattle, WA 98122, USA}\\
{\it $ ^c)$  Institut fuer Theoretische Physik,
  Universitaet Giessen,
  35392 Giessen, Germany (present address)}\\
{\it $ ^d)$ Institute for Nuclear Theory, Box 351550, University of Washington, Seattle, WA 98195, USA}}
\maketitle
\begin{abstract}
We use the 
meson cloud model and the Sullivan mechanism to estimate the sea flavor 
asymmetry in the $\Sigma^+$ baryon and calculate the distribution functions of
both sea and valence quarks. We find large deviations from SU(3).
\end{abstract}
\newpage
\section {Introduction} 
    By now, there have been a number of experiments which point to a flavor 
asymmetry in the sea of the proton \cite{E772,NA51,E866}. Although the 
exclusion principle \cite{Thom} and charge asymmetry \cite{Kumano,Thomas} 
may contribute to the observed asymmetries, they 
are not sufficiently large to explain the data \cite{Thomas}. 
The two most noteworthy experiments are the deviation of the Gottfried
sum rule from 1/3, i.e., $S_G = 0. 235\pm 0.026$ \cite{NMC} and the recent 
Drell-Yan measurements
of $\bar{d}/\bar{u}$ by E866 \cite{E866} and NA51 \cite{NA51}. The
CERN measurement
gives $\bar{u}/\bar{d} \simeq 0.51$ at $x=0.18$ and the Fermilab one 
obtains the major part of the
$x$-distribution  of $\bar{d}(x) - \bar{u}(x)$
, finding $\bar{d}/\bar{u} \simeq 1.5$ at $x \simeq 0.2$. The most 
reasonable explanation of these large deviations from unity or 1/3 is that
there is a flavor asymmetry in the sea quark distributions; this flavor
asymmetry is readily understood if a meson cloud surrounds the quarks, in 
that a proton can change into a neutron by emitting a $\pi^+ (u \bar{d})$,
as first pointed out by Thomas \cite{Thom} and later by Henley and
Miller \cite{HM}. The 
Sullivan process allows one to calculate the $x$-distribution of the 
asymmetry, as has been done by Speth and collaborators 
\cite{Sp} and others (see \cite {Sp} for references); reasonable agreement with
experiment is obtained. \\

    More recently, Alberg et al. \cite{Alb} have pointed out that a
test of the 
meson cloud model is feasible by carrying out Drell-Yan experiments with
$\Sigma$ beams on protons and deuterium. In this model, the flavor
asymmetry in the $\Sigma^+$, for instance, 
is expected to be even larger than in
the proton ($p$), so that $\bar{d}/\bar{u}$ should be much larger than
1. This is in contrast to the prediction of SU(3), under which
$p\rightarrow \Sigma^+$ by $d(\bar{d})\leftrightarrow s(\bar{s})$, from which it follows 
that $\bar{d}/\bar{u} < 1$ in the $\Sigma^+$.\\

In this paper we use the convolution method with the Sullivan process to
calculate the valence and sea quark distributions of the $\Sigma^+$ in 
the meson cloud model.

\section{Methodology}

    In the convolution method, the physical $\Sigma^+$ wave function is composed 
of the following Fock states
\begin{equation}
\vert\Sigma^+\rangle= \sqrt{Z}(\vert\Sigma^+\rangle_{bare}+\sum_{MB}
\int dyd^2\vec{k}_\perp\phi_{BM}(y,\vec{k}_\perp) 
\vert B(y,\vec{k}_\perp);M(1-y,-\vec{k}_\perp)\rangle) \;,
\end{equation}
where $\phi_{BM}(y,\vec{k}_\perp)$ is the probability amplitude to find a 
physical $\Sigma^+$ in a state consisting of a 
virtual baryon $B$ and a virtual meson $M$ with longitudinal momentum 
fractions $y$ and $1-y$, 
and transverse momenta $\vec{k}_\perp=(k_\perp
\cos(\varphi),k_\perp\sin(\varphi))$ and $-\vec{k}_\perp$, respectively. 
The wave function renormalization factor $Z$ is a measure of
the probability of finding the "bare" $\Sigma^+$,
that is the $\Sigma^+$ without a meson cloud in the physical $\Sigma^+$.\\

We assume that the $\Sigma^+(uus)$ will have components $\Lambda^0(uds)\pi^+
(u\overline{d})$,\\ $\Sigma^0(uds)\pi^+(u\overline{d})$, 
$\Sigma^+(uus)\pi^0(\frac{1}{\sqrt{2}}[d\overline{d}-u\overline{u}])$ 
and $p(uud)\overline{K}^0(\overline{d}s)$ \cite{Alb}. We neglect higher 
mass components.\\

In the infinite momentum frame (IMF, $\vert\vec{p}\vert\to\infty$ with 
$\vec{p}$ the $\Sigma^+$ momentum) the contribution 
of a certain Fock state, $BM$, to the $\Sigma^+$ quark distribution can be 
written in terms of its quark components as
\begin{equation}
\delta q_{\Sigma^+}(x)=\sum_{MB}(\int_x^1f_{MB/\Sigma^+}(y)q_M(\frac{x}{y})
\frac{dy}{y}+\int_x^1f_{BM/\Sigma^+}(y)q_B(\frac{x}{y})\frac{dy}{y}) \; ,
\end{equation}
where the splitting functions $f_{MB/\Sigma^+}(y)$ and $f_{BM/\Sigma^+}(y)$ 
are related to the probability amplitude $\phi_{BM}$ in the IMF via
\begin{equation}
f_{BM/\Sigma^+}(y)=\int_0^\infty dk_\perp^2\vert\phi_{BM}(y,k_\perp^2)
\vert^2\; ,
\end{equation} 
\begin{equation}
f_{MB/\Sigma^+}(y)=\int_0^\infty dk_\perp^2\vert\phi_{BM}(1-y,k_\perp^2)
\vert^2.
\end{equation}
In terms of these splitting functions the wave function renormalization 
constant Z is given by
\begin{equation} 
Z=[1+\sum_{BM}\langle f_{BM/\Sigma^+}\rangle]^{-1}\equiv [1+\sum_{BM}\int_0^1f_{BM/
\Sigma^+}(y)dy \;]^{-1} ,
\end{equation} 
and the quark distribution functions $q_{\Sigma^+}$ of a $\Sigma^+$ within 
the Fock state expansion are given as
\begin{equation}
q_{\Sigma^+}(x)=Z(q_{\Sigma^+}^{bare}(x)+\delta q_{\Sigma^+}(x))
\end{equation}
where $q_{\Sigma^+}^{bare}$ is the quark distribution of the bare $\Sigma^+$.\\

The next step is to calculate the splitting functions $f_{MB/\Sigma^+}$ and 
$f_{BM/\Sigma^+}$. We do this using time ordered perturbation theory (TOPT) 
in the IMF, following the steps of reference \cite{Sp}. In TOPT in the IMF 
one can write the probability amplitudes $\phi_{BM}(y,k_\perp^2)$ explicitly 
as
\begin{equation}
\phi_{BM}(y,k_\perp^2)=\frac{\sqrt{m_{\Sigma^+}+m_B}V_{IMF}(y,k_\perp^2)}
{2\pi\sqrt{y(1-y)}(m_{\Sigma^+}^2-M_{BM}^2(y,k_\perp^2))}
\end{equation}
where $M_{BM}^2(y,k_\perp^2)$ is the invariant mass squared of the 
intermediate $BM$ Fock state
\begin{equation}
M_{BM}^2(y,k_\perp^2)=\frac{m_B^2+k_\perp^2}{y}+\frac{m_M^2+k_\perp^2}{1-y}
\end{equation}
and $V_{IMF}$ denotes the vertex function in the IMF-limit. Vertices
involving point-like particles automatically fulfill the symmetry relation
\begin{equation}
f_{MB/\Sigma^+}(y)=f_{BM/\Sigma^+}(1-y)
\end{equation}
but since hadrons have an extended structure one has to introduce 
phenomenological vertex form factors which parameterize the unknown 
microscopic effects. Therefore the vertex function $V(y,k_\perp^2)$ is 
replaced by $G(y,k_\perp^2)$$V(y,k_\perp^2)$ and from equation (9) we get 
the restriction
\begin{equation}
G_{BM}(y,k_\perp^2)=G_{MB}(1-y,k_\perp^2) \; .
\end{equation}

This is satisfied by the exponential form \cite{Sp}
\begin{equation}
G_{\Sigma^+BM}(y,k_\perp^2)=\exp(\frac{m_{\Sigma^+}^2-M_{BM}^2(y,k_\perp^2)}
{2\Lambda^2}) \; ,
\end{equation}
where $\Lambda$ is a cut-off parameter  
which we use in our further calculations.\\

Equations (3) and (7) allow us to write the splitting functions as
\begin{equation}
f_{BM/\Sigma^+}(y,k_\perp^2)=\frac{1}{4\pi^2}\frac{m_{\Sigma^+}m_B}{y(1-y)}
\frac{\vert V_{IMF}\vert^2}{[m_{\Sigma^+}^2-M_{BM}^2(y,k_\perp^2)]^2}.
\end{equation}
One gets the spin averaged vertex functions in the IMF from the interaction 
Lagrangian density
$\mathcal L\mathit=ig\overline{\phi}\gamma_5\pi\phi$ where $\phi$ 
denotes a baryon $(\Lambda^0$, $\Sigma^0$, $\Sigma^+$, $p)$ 
and $\pi$ a pseudo scalar field $(\pi^+$, $\pi^0$, $\overline{K}^0)$ \cite{Holt}. 
For the vertex 
\begin{equation}
\Sigma^+(helicity=1/2)\to baryon(helicity=\lambda)+meson(helicity=\lambda')\;,
\end{equation}
the vertex function $V_{IMF}^{\lambda\lambda'}(y,k_\perp^2)$ is given by
\begin{equation}
V_{IMF}^{\frac{1}{2}0}=\frac{g}{2}\frac{ym_{\Sigma^+}-m_B}{\sqrt{ym_{\Sigma^+}
m_B}},
\end{equation}
\begin{equation}
V_{IMF}^{-\frac{1}{2}0}=\frac{g}{2}e^{-i\varphi}\frac{k_\perp}{\sqrt{ym_
{\Sigma^+}m_B}}\;,
\end{equation}
and when spin averaged we obtain
\begin{equation}
\vert V_{IMF}\vert^2=\frac{g^2}{4}\frac{(ym_{\Sigma^+}-m_B)^2+k_\perp^2}
{ym_{\Sigma^+}m_B}\;.
\end{equation}
Hence, in our case the splitting function is given by
\begin{equation}
f_{BM/\Sigma^+}(y)=\frac{g^2}{16\pi^2}\frac{1}{y^2(1-y)}\int_0^\infty dk_
\perp^2\vert G_{\Sigma^+BM}(y,k_\perp^2)\vert^2\frac{(ym_{\Sigma^+}-m_B)^2
+k_\perp^2}{[m_{\Sigma^+}^2-M_{BM}^2(y,k_\perp^2)]^2}.
\end{equation}
Using the following coupling constants g \cite{DKP}\\
\begin{tabbing}
$\frac{1}{4\pi}g_{\Lambda^0\pi^+/\Sigma^+}^2$ \= $=11.8$\\
$\frac{1}{4\pi}g_{\Sigma^0\pi^+/\Sigma^+}^2$ \> $=13.0$\\
$\frac{1}{4\pi}g_{\Sigma^+\pi^0/\Sigma^+}^2$ \> $=13.0$\\
$\frac{1}{4\pi}g_{p\overline{K}^0/\Sigma^+}^2$ \> $=2.0$\\
\end{tabbing}
and the cutoff parameter $\Lambda=1.08$ GeV
for most of our work, \cite{Sp}
we just 
need the quark distributions in the bare particles 
as an additional input.\\

For some of the following, we
use Holtmann's parameterization of the quark distribution function in the 
bare nucleon \cite{HDiss} in which he assumes a symmetric sea, $\bar{Q}_{bare}^p$
\begin{equation}
xu_{v,bare}^p(x)=0.62x^{0.37}(1-x)^{2.5}(1+11x),
\end{equation}
\begin{equation}
xd_{v,bare}^p(x)=0.04x^{0.10}(1-x)^{4.7}(1+102x),
\end{equation}
\begin{equation}
x\bar{Q}_{bare}^p=0.11(1-x)^{15.8},
\end{equation}
\begin{equation}
\bar{Q}_{bare}^p=u_{sea,bare}^p=\overline{u}_{sea,bare}^p=d_{sea,bare}^p=
\overline{d}_{sea,bare}^p=2s_{sea,bare}^p=2\overline{s}_{sea,bare}^p \; .
\end{equation}
In addition to this form for $\bar{Q}_{bare}^p$, we also take a form which
is tied to the 
recent determination of the gluon distribution \cite {Vogt},
\begin{equation}
x\bar {Q}_{bare}'^{p} = 0.0124 x^{-0.36} (1-x)^{3.8} \; .
\end{equation}
Since the gluon splits into the "perturbative" or bare sea quarks,  this
distribution should be close to that of those quarks. 
The advantage of this choice will become clear in the next section.
For the quark distribution in the pion \cite{Sut} we take 
\begin{equation}
xq_v(x)=0.99x^{0.61}(1-x)^{1.02},
\end{equation}
\begin{equation}
xq_{sea}(x)=0.2(1-x)^{5.0}\; ,
\end{equation}
where 20\% of the pion's momentum is assumed to be carried by the 
symmetrical sea, which is presumed to be due to gluon splitting.\\

To begin with, we determined
the quark distributions of the bare hyperons 
by using SU(3) symmetry for the valence quarks; that is we neglected the 
mass difference between the $s$ and $u$ and $d$ quarks. We do not show the 
results because we do not believe that this is a realistic choice.
Look at the $\overline{K}^0$. Experiments \cite{Bad} show that 
\begin{equation}
\frac{\overline{u}^{K^-}}{\overline{u}^{\pi^-}}\sim(1-x)^{0.18\pm 0.07}.
\end{equation}
Using this we can take the parametrization for the quark distribution
in the pion to get the $\overline{u}$ distribution in the $K^-$ and
through charge independence the $\overline{d}$ distribution in the 
$\overline{K}^0$. To get the s distribution in the $\overline{K}^0$ we
assume that the gluon and the light sea quark distributions in the kaon
and the pion are the same and hence carry the same momentum fraction
in both particles. We also assume the following form of the s quark 
distribution
\begin{equation}
xs_v(x)\sim x^{0.61}(1-x)^a
\end{equation}
where the parameter $a$ is determined by matching up the right momentum fraction. The valence quark distributions in the $\overline{K}^0$ are then given by
\begin{equation}
x\overline{d}_v(x)=1.05x^{0.61}(1-x)^{1.20},
\end{equation}
\begin{equation}
xs_v(x)=0.94x^{0.61}(1-x)^{0.86}.
\end{equation}
Compared to
the pion, where $d(x)$ and $\overline{u}(x)$ look the same, the s
distribution in the $\overline{K}^0$ now peaks at higher $x$ than the 
$\overline{d}$ distribution, which reflects the fact that the s quark
is heavier than the u and the d.\\

Our starting point for the bare $\Sigma^+$ is Holtmann's
parametrization for the bare nucleon, equations 18-21. Again we
assume only a change in the $(1-x)$-part of the parametrizations and
that the light sea quarks and gluons in the bare $\Sigma^+$ and the bare nucleon look the same. To account for the higher mass of the s quark we then make the Ansatz
\begin{equation}
\frac{\int_0^1xu_{v,bare}^{\Sigma^+}(x)dx}{\int_0^1xs_{v,bare}^{\Sigma^+}(x)dx}=\frac{\int_0^1xu_{v,bare}^p(x)dx}{\int_0^1xd_{v,bare}^p(x)dx}\cdot\frac{m_{d,con}}{m_{s,con}}
\end{equation}
where we take model dependent constituent quark masses to get
\begin{equation}
\frac{m_{d,con}}{m_{s,con}}\simeq\frac{336MeV}{540MeV}
\end{equation}
The quark distributions in the bare $\Sigma^+$ are then 
\begin{equation}
xu_{v,bare}^{\Sigma^+}(x)=0.82x^{0.37}(1-x)^{3.89}(1+11x),
\end{equation}
\begin{equation}
xs_{v,bare}^{\Sigma^+}(x)=0.03x^{0.1}(1-x)^{1.76}(1+102x).
\end{equation}
\begin{equation}
x\bar{Q}_{bare}^{\Sigma^+}=x\bar{Q}_{bare}^{p}
\end{equation}
We can also determine the distributions in the bare $\Sigma^0$ and
$\Lambda^0$ via charge independence and SU(3):
\begin{tabbing}
$\Sigma^0$: \= $\bar{Q}_{bare}^{\Sigma^0}=\bar{Q}_{bare}^{\Sigma^+}$ \hspace{5cm} \= $\Lambda^0$: \= $\bar{Q}_{bare}^{\Lambda^0}=\bar{Q}_{bare}^{\Sigma^+}$\\
\> $u_{v,bare}^{\Sigma^0}=\frac{1}{2}u_{v,bare}^{\Sigma^+}$ \> \> $u_{v,bare}^{\Lambda^0}=\frac{1}{2}u_{v,bare}^{\Sigma+}$\\
\> $d_{v,bare}^{\Sigma^0}=\frac{1}{2}u_{v,bare}^{\Sigma+}$ \> \> $d_{v,bare}^{\Lambda^0}=\frac{1}{2}u_{v,bare}^{\Sigma^+}$\\
\> $s_{v,bare}^{\Sigma^0}=s_{v,bare}^{\Sigma^+}$ \> \> $s_{v,bare}^{\Lambda^0}=s_{v,bare}^{\Sigma^+}$.\\
\end{tabbing}
Since we assume no change in the symmetric sea of the bare particles, the only 
variance in the antiquark distributions of the physical $\Sigma^+$
comes from the different parametrization of the $\overline{K}^0$ which
only affects $\overline{d}(x)$. There is also a change in $d(x)$ due
to the more realistic $s$ quark distributions for the $\Sigma^0$ and the
$\Lambda^0$. The splitting
$\Sigma^+\to p\overline{K}^0$ is nearly negligible. Hence taking into
account the larger s quark mass does not noticeably change our results for
$\bar{d}/\bar{u}$ and the difference $\bar{d} - \bar{u}$ 
from those using SU(3) parameters. However, the valence quark distributions 
in the physical $\Sigma^+$ do change.

\section {Results}

We first test our model for the measured sea quark distributions in the proton.
We show both $(\bar{d} - \bar{u})$ and $\bar{d}/\bar{u}$ in our model with 
no $\Delta$ and no mesons more massive than the pion in Figs. 1 and 2. 
The comparison with the E866 data shows that $(\bar{d} - \bar{u})$ agrees 
reasonably with experiment, but that the ratio $\bar{d}/\bar{u}$ does not 
turn over towards 1 at higher value of $x$
with the parameterization of the bare sea used by Holtmann\cite {HDiss}.
This problem is also found for other meson cloud and chiral models, as
has been recently noted by Peng et al.\cite{Peng}. However, we find
that the ratio does turn over
(although not sufficiently fast) for $\bar{Q}_{bare}'^p$ given by Eq. 22. 
It is true
that the inclusion of the $\Delta$ would help, but we believe that the 
splitting of the gluon into $ q \bar{q}$ pairs is the dominant cause of 
the return of the ratio $ \bar{d}/\bar{u}$ towards unity at $x > 0.3$ .
We believe that a parameterization for $\bar{Q}_{bare}^p$ can be found that
is consistent with the gluonic data and with the ratio $\bar{d}/\bar{u}$.
For a further discussion, see also \cite{Mel}.\\

The four splitting functions for the $\Sigma^+$
 are shown in Fig. 3; it is clear that the contribution from the 
$ p \bar{K}^0$ state is very small. In Fig. 4  we show the calculated
valence quark momentum distributions. The s quark distribution peaks at a slightly higher
value of $x$ than that of the $u$ quark momentum distribution 
due to its larger mass.  The momentum distribution of the sea quarks is shown
in Figs. 5-7. We show the distributions for both the Holtmann \cite {HDiss}
and the gluonic \cite {Vogt} bare $\Sigma^+$ distributions. The $x d$ and 
$x \bar{d}$ distributions are slightly different, but the difference is so 
small that we do not attempt to show it. The difference $(\bar{d} -
\bar{u})$ shown in Fig. 8 is, of course, independent of the bare $\Sigma^+$'s
quark distribution, since it is due to 
gluon splitting. It is interesting to see in Fig. 9
how much $\bar{r}_\Sigma \equiv \bar{d}/\bar{u}$ distribution depends on that
of the bare sea. In both cases considered the ratio is larger than $\bar{r}
_p$ in the proton. For the Holtmann \cite {HDiss} bare quark distribution 
$\bar{r}_\Sigma$ increases with increasing $x$ and does not turn over.
For the gluonic distribution, on the other hand, $\bar {r}_\Sigma$ does not 
rise half as high and approaches 1 as $x \rightarrow 0.6$. We show 
$(\bar{d} - \bar{u})_\Sigma/ (\bar{d}-\bar{u})_p$ in Fig. 10 and compare
$\bar{r}_\Sigma$ to $\bar{r}_p$ in Fig. 11. In the latter figure we note that
$\bar{r}_\Sigma / \bar{r}_p$ falls below 1 for the gluonic-like distribution 
at high $x$, but the distribution functions are very small here. \\

We also tested the influence of the cut-off parameter $\Lambda$ on our 
results. Since a higher cut-off primarily makes the amplitudes of the 
splitting functions larger but does not change their shape very much, 
the sea quarks will carry more of the $\Sigma^+$ momentum while the momentum 
fraction carried by the valence s quarks stays the same and that of the 
valence u quarks gets smaller. The ratio $\overline{d}/\overline{u}$ gets 
smaller and the difference $\overline{d}-\overline{u}$ gets larger with
growing $\Lambda$ [ at all x]. 
\section {Conclusions}
In conclusion, we have used the meson cloud model to calculate the
valence and sea quark distributions in the $\Sigma^+$. The same model
can of course be used to calculate the valence and sea quark
distributions in the $\Sigma^0, \Lambda^0$ and $\Sigma^-$. Now that 
quark distributions in the proton have been well-determined, it  
is possible to use $\Sigma^+$  and $\Sigma^-$
beams and inclusive Drell-Yan reactions
on protons, alone, to obtain similar information for the hyperons. 
Nuclear binding corrections for deuterium targets are thereby
avoided. For instance
if $x(p)$ is large and $x(\Sigma^\pm)$ is small, the measurement
\begin{equation}
\sigma(\Sigma^+p) - \sigma(\Sigma^-p) \propto(\bar{d}^\Sigma -\bar{u}^\Sigma)
(\frac{1}{9}d^p - \frac{4}{9}u^p)\
\end{equation}
is sensitive to $(\bar{d} - \bar{u})$ in the $\Sigma^+$.

\section {Acknowledgments}

We wish to thank J.C. Peng for fruitful discussions and for providing
helpful information. This work has been
supported in part by the U. S. Department of Energy, Contract No.
DE-FG03-97ER41014, and by the National
Science Foundation, Grant No. PHY-97-22023.

\newpage

\begin{figure}
\epsfxsize=10cm
\hfil\epsfbox{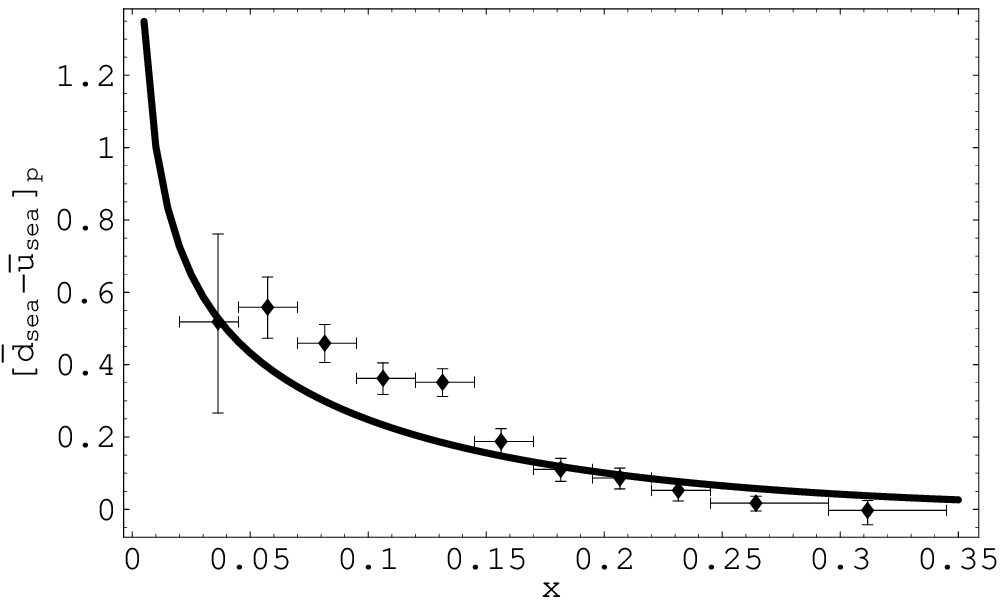}\hfill
\caption{Comparison of our meson model with data [3] for $(\bar{d} -
\bar{u})$.							}
\label{1}
\end{figure}

\newpage

\begin{figure}
\epsfxsize=10cm
\hfil\epsfbox{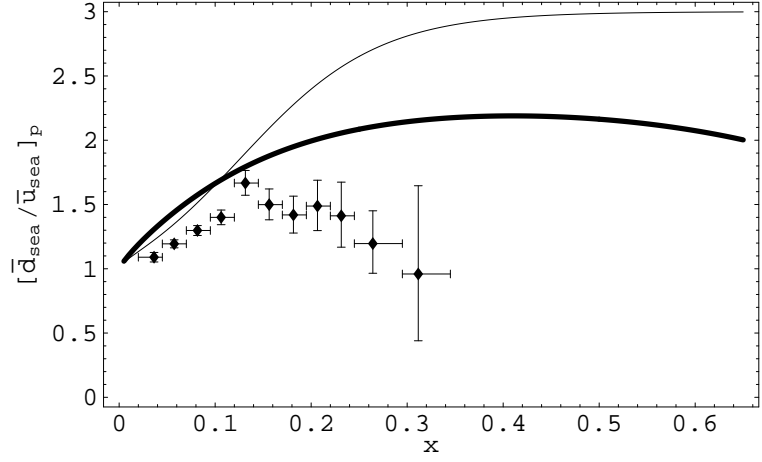}\hfill
\caption{Comparison of our model with data for $\bar{d}/\bar{u}$ and
data [3]. The light line is for the bare distribution of Eq.(20) and
the dark line for Eq.(22).}
\label{2}
\end{figure}
\newpage

\begin{figure}
\epsfxsize=10cm
\hfil\epsfbox{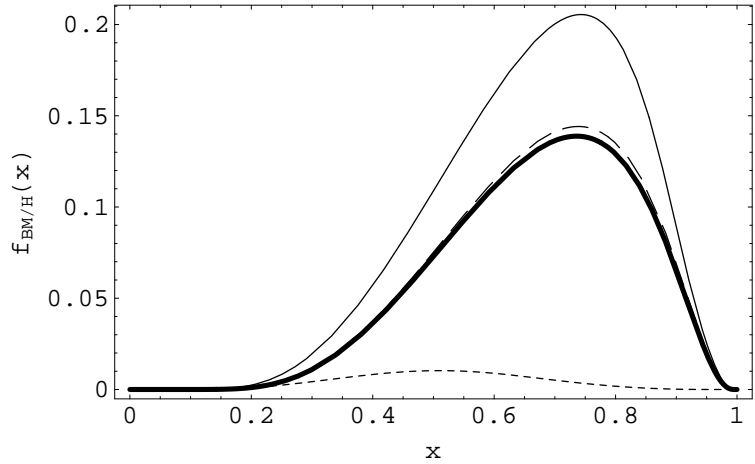}\hfill
\caption{ The splitting functions for $\bar{K}p$ (short dashes), 
	  $\Sigma^+ \pi^0$
          (long dashes), $\Sigma^0 \pi^+$ (heavy line) and $\Lambda^0
          \pi^+$ (light line).}
\label{3}
\end{figure}
\newpage

\begin{figure}
\epsfxsize=10cm
\hfil\epsfbox{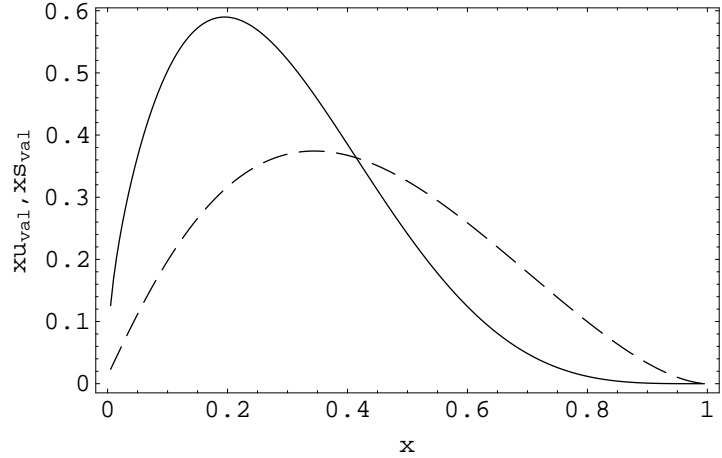}\hfill
\caption{Valence quark distributions in the $\Sigma^+$. The solid line is that
          of the $u$ quarks and the dashed line is that for $s$ quarks.}
\label{4}
\end{figure}
\newpage

\begin{figure}
\epsfxsize=10cm
\hfil\epsfbox{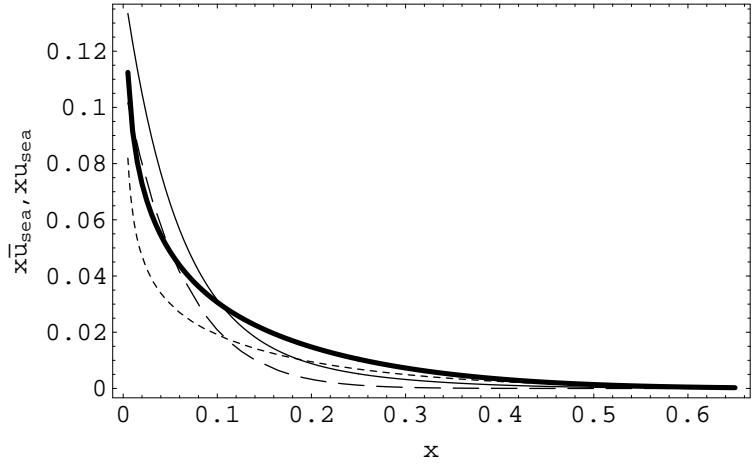}\hfill
\caption{ The sea $u$ and $\bar{u}$ quark momentum distributions. The light
          line and long dashes are for Eq.(20) and the heavy line and short
          dashes for Eq.(22). The dashed lines are the bare distibutions and 
          the solid ones for the physical $\Sigma^+$.}
\label{5}
\end{figure}
\newpage

\begin{figure}
\epsfxsize=10cm
\hfil\epsfbox{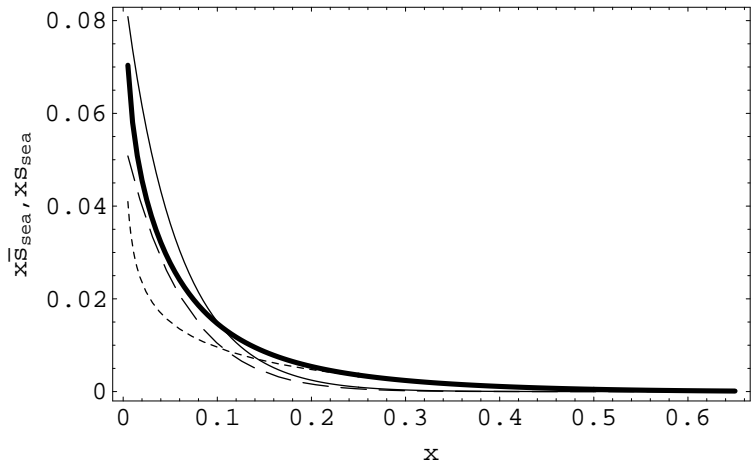}\hfill
\caption{ The sea $s$ and $\bar{s}$ quark momentum distributions. See Fig. 5
          for details.}
\label{6}
\end{figure}
\newpage

\begin{figure}
\epsfxsize=10cm
\hfil\epsfbox{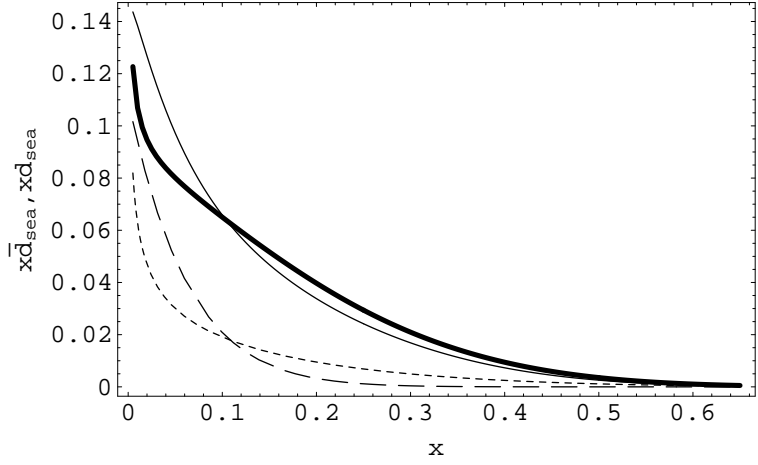}\hfill
\caption{The sea $d$ and $\bar{d}$ quark momentum distributions. See Fig. 5
          for details. Although the distributions differ slightly for $d$ and
          $\bar{d}$, the differences are small and not shown.}
\label{7}
\end{figure}
\newpage

\begin{figure}
\epsfxsize=10cm
\hfil\epsfbox{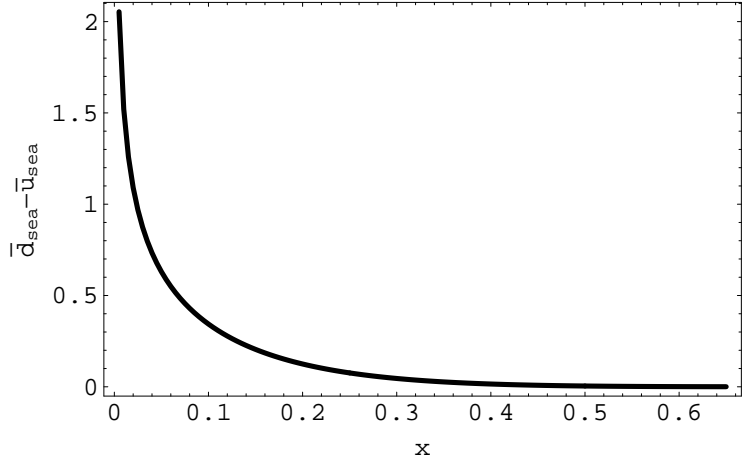}\hfill
\caption{The difference $(\bar{d} - \bar{u})$ for the $\Sigma^+$.}
\label{8}
\end{figure}
\newpage

\begin{figure}
\epsfxsize=10cm
\hfil\epsfbox{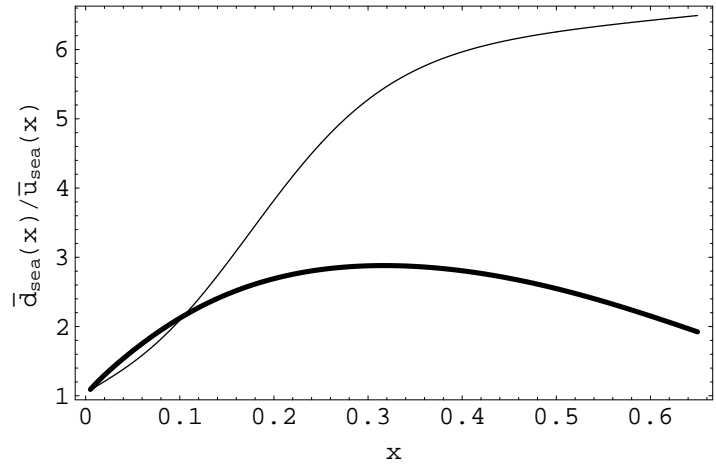}\hfill
\caption{The ratio $\bar{d}/\bar{u}$ for the $\Sigma^+$.}
\label{9}
\end{figure}
\newpage

\begin{figure}
\epsfxsize=10cm
\hfil\epsfbox{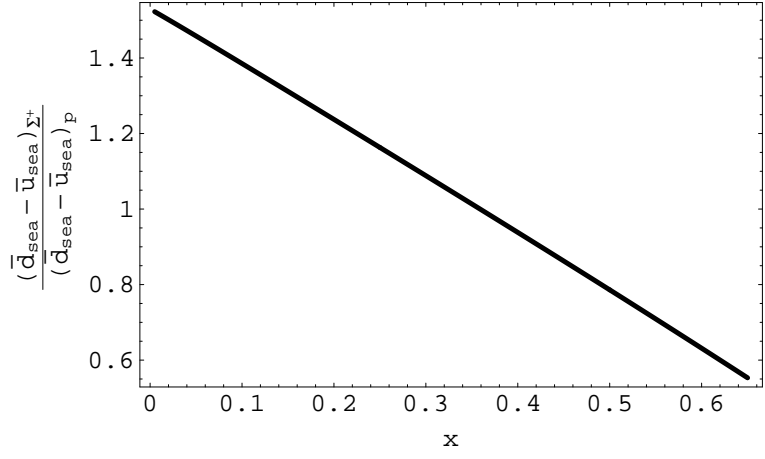}\hfill
\caption{The ratio $\frac{(\bar{d} - \bar{u})_{\Sigma^+}}{(\bar{d} - \bar{u}
          )_p}$.}
\label{10}
\end{figure}
\newpage

\begin{figure}
\epsfxsize=10cm
\hfil\epsfbox{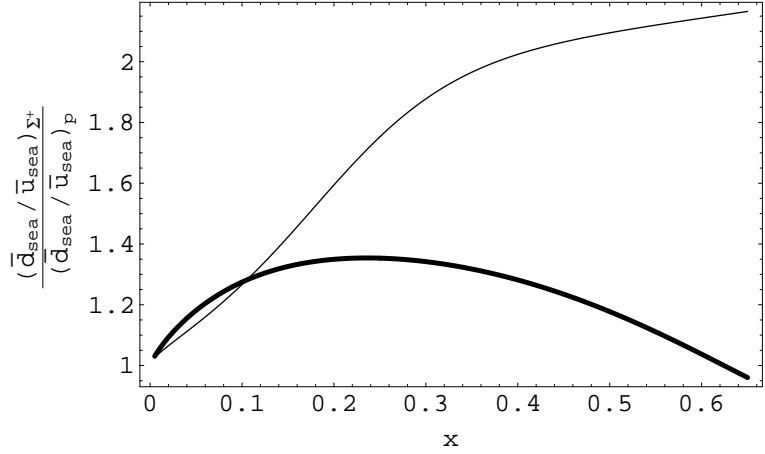}\hfill
\caption{The ratio $\bar{r}_{\Sigma^+}/\bar{r}_p$. The light line is for
          Eq.(20) and the heavy one for Eq. (22).}
\label{11}
\end{figure}

\end{document}